\documentclass[useAMS,usenatbib]{mn2e}
\usepackage{graphicx,amssymb, epstopdf}

\newcommand{\vc}[1]{\bmath{#1}}

\newcommand{\vB}{\vc{B}}
\newcommand{\vb}{\vc{b}}
\newcommand{\vg}{\vc{g}}
\newcommand{\vu}{\vc{u}}
\newcommand{\dl}{\rmn{d}}

\newcommand{\beq}{\begin{equation}}
\newcommand{\eeq}{\end{equation}}
\newcommand{\bea}{\begin{eqnarray}}
\newcommand{\eea}{\end{eqnarray}}
\newcommand{\lt}{\left}
\newcommand{\rt}{\right}

\title[Local heat flux in an intracluster medium]{Suppression 
of local heat flux in a turbulent magnetized intracluster medium}
\author[S. V. Komarov, E. M. Churazov, A. A. Schekochihin and J. A. ZuHone]{
S. V. Komarov$^{1,2,3}$, E. M. Churazov$^{1,2}$, 
A. A. Schekochihin$^{4}$ and J. A. ZuHone$^{5}$\\
$^{1}$Max Planck Institute for Astrophysics, Karl-Schwarzschild-Strasse 1, 
85741 Garching, Germany\\
$^{2}$Space Research Institute (IKI), Profsouznaya 84/32, Moscow 117997, Russia\\
$^{3}$Moscow Institute of Physics and Technology (MIPT), Institutsky pereulok 9, 
Dolgoprudny 141700, Moscow Region, Russia\\
$^{4}$The Rudolf Peierls Centre for Theoretical Physics, University of Oxford, 
1 Keble Road, Oxford OX1 3NP, United Kingdom\\
$^{5}$Astrophysics Science Division, Laboratory for High Energy Astrophysics, 
Code 662, NASA/Goddard Space Flight Center, Greenbelt, \\MD 20771, USA} 
\begin{document}
\maketitle
\label{firstpage}

\begin{abstract}
X-ray observations of hot gas in galaxy clusters often show steeper
temperature gradients across cold fronts -- contact discontinuities,
driven by the differential gas motions. These sharp (a few kpc wide)
surface brightness/temperature discontinuities would be quickly
smeared out by the electron thermal conduction in unmagnetized plasma,
suggesting significant suppression of the heat flow across the
discontinuities. In fact, the character of the gas flow near cold 
fronts is favorable for suppression of conduction by aligning magnetic 
field lines along the discontinuities. We argue that a similar mechanism 
is operating in the bulk of the gas. Generic 3D random isotropic and
incompressible motions increase the temperature gradients 
(in some places) and at the same time suppress the {\it local} conduction 
by aligning the magnetic field lines perpendicular to the {\it local} 
temperature gradient. We show that the suppression of the effective 
conductivity in the bulk of the gas can be linked to the increase of 
the frozen magnetic field energy density. On average the rate of decay 
of the temperature fluctuations~$\dl \langle \delta T^2 \rangle / \dl t$  
decreases as $\langle B^2 \rangle ^{-1/5}$.

\end{abstract}


\section{Introduction}
X-ray observations of galaxy clusters reveal significant spatial
fluctuations of the gas temperature in a range of spatial scales 
\citep[e.g.][]{Markevitch2003}. Given a temperature map with prominent
fluctuations, it is possible to calculate an upper limit on the effective
thermal conductivity, provided that the lifetime of the fluctuations
can be estimated. It turns out to be at least an order of magnitude
lower than the Spitzer conductivity for unmagnetized plasma 
\citep{Ettori2000, Markevitch2003}.

Heat conduction in the intracluster medium (ICM) is primarily along 
the field lines because the Larmor radius of the particles is very small 
compared to the collisional mean free path \citep{Braginskii1965}. 
The ICM undergoes turbulent motion in a range of spatial scales
\citep{Inogamov2003, Schuecker2004, Schek2006turb, Subramanian2006, 
Zhuravleva2011}. As the magnetic field is, to a good approximation, 
frozen into the ICM, the field lines become tangled by gas motions and 
their topology changes constantly. Four main effects should be considered. First, 
parallel thermal conduction along stochastic magnetic field lines may be
reduced because the heat-conducting electrons become trapped and
detrapped between regions of strong magnetic field (magnetic mirrors;
see \citealt{CC1998, CC1999, MalyshkinKulsrud2001, Albright2001}). 
Secondly, diffusion in the transverse direction may be boosted due to 
spatial divergence of the field lines \citep{Skilling1974, RR1978, CC1998, 
Narayan2001, ChandranMaron2004}. Thirdly, there is effective diffusion 
due to temporal change in the magnetic field (`field-line wandering'). 
Finally, if one is interested in temperature fluctuations and their 
diffusion, one must be mindful of the fact that the temporal evolution 
of the magnetic field is correlated with the evolution of the temperature 
field because the field lines and the temperature are advected by the same 
turbulent velocity field.

In this paper, we focus on the last effect. The more conventional 
approach, often used to estimate the relaxation of the temperature 
gradients, is to consider the temperature distribution as given and 
study the effect of a tangled magnetic field on the heat conduction. 
However, the direction and value of the fluctuating temperature
gradients are not statistically independent of the direction of the
magnetic-field lines because the latter are also correlated with the 
turbulent motions of the medium. We argue that, dynamically, the
fluctuating gradients tend to be oriented perpendicular to the field
lines and so heat fluxes are the more heavily suppressed the stronger
the thermal gradients are. We also establish the relationship between 
the average conductivity and the growth of the magnetic energy density. 

The structure of the paper is as follows. In Section~\ref{sec:qual_pic}, we 
provide a qualitative explanation of the correlation between the 
temperature gradients and the magnetic-field direction, accompanied by 
a number of numerical examples. In Section~\ref{sec:results}, a theoretical 
framework for modelling this effect is presented and the joint PDF of 
the thermal gradients, the angles between these gradients and the 
magnetic-field lines and the magnetic-field strength is derived in the solvable 
case of a simple model velocity field. The connection between the effective 
conductivity and the increase of the magnetic energy density is established. 
Analytical results are supplemented by numerical calculations in 
Section~\ref{sec:fin_corr}, which extrapolate our results to the case of a 
more general velocity field. In Section~\ref{sec:disc}, we discuss the 
assumptions that have been necessary to enable analytical treatment, 
the consequent limitations on the applicability of our results, 
and also present some numerical tests using a global dynamical 
cluster simulation, which suggest that, at least qualitatively, our 
results survive when most of the simplifying assumptions are relaxed. 
Finally, in Section~\ref{sec:concl}, we sum up our findings.

\section{Qualitative discussion}
\label{sec:qual_pic}
We consider a volume of plasma with high electric conductivity
and frozen-in magnetic field tangled on a scale much greater than 
the mean free path of the particles. We also assume the plasma motions 
to be incompressible, which is a good approximation for subsonic dynamics. 
Across the paper we treat the temperature as a passive scalar.

\subsection{ Illustrative example: conduction between 
converging layers of magnetised plasma}
\label{sec:example}

\begin{figure}
\includegraphics[width=84mm]{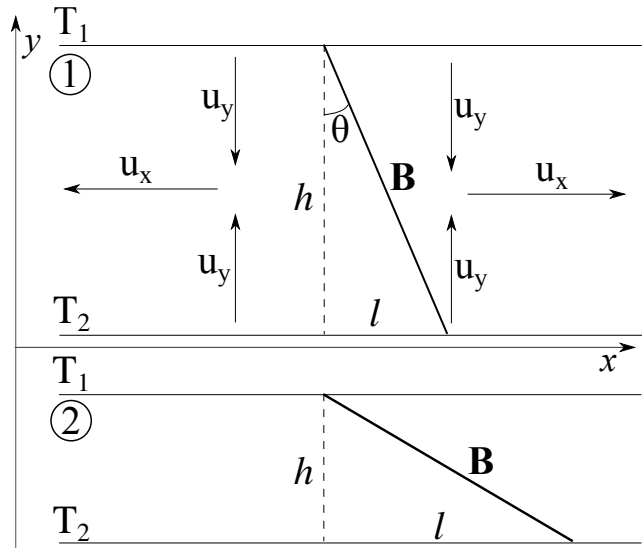}
\caption{Correlated changes of the temperature gradients and the 
inclination of the magnetic-field lines in the case of a converging 
incompressible flow: plane parallel layers at different temperatures. 
Converging flow with $\partial _y u_y <0$ reduces $h$ and increases 
the temperature gradient $(T_2 - T_1)/h$, but suppresses heat flux. 
The solid line represents the direction of the 
magnetic field. If the medium is incompressible then $l \times h$ is
conserved (in the absence of tangential shears).}
\label{fig:conv_flow}
\end{figure}

Consider two parallel layers of an incompressible medium vertically 
separated by distance $h$ with temperatures $T_1 \neq T_2$. This is 
illustrated in Fig.~\ref{fig:conv_flow}: the direction of the field 
line is shown with the inclined solid line, making an angle $\theta$ 
with the vertical, so $\cos{\theta}=h/\sqrt{h^2+l^2}$, where $l$ is 
the horizontal distance between the footpoints of the field line anchored 
in the two layers. An incompressible flow with $\partial _y u_y <0$ 
reduces $h$ and increases $l$ so that $l \times h$ is conserved (in the 
absence of tangential shear). Here we are interested in the heat exchange 
between the layers, i.e. only the component of the heat flux along the 
temperature gradient $Q_{\nabla T}$ has to be calculated:
\beq
Q_{\nabla T} = \chi(\vb\cdot\nabla T)\cos{\theta} =
\chi \frac{T_2 - T_1}{\sqrt{h^2 + l^2}} \frac{h}{\sqrt{h^2 + l^2}}.
\eeq
Let $h(t)=h_0 f(t)$ and $l(t)=l_0/f(t)$. Then
\beq
Q_{\nabla T}=\chi \frac{T_2 - T_1}{h_0} \frac{f}{\displaystyle f^2 + 
(l_0/h_0)^2 f^{-2}},
\eeq
where $\chi$ is the parallel thermal diffusivity coefficient 
\citep{Braginskii1965}, which is assumed constant across the volume for 
simplicity. Therefore, in the limit of $f \rightarrow 0$, $Q_{\nabla T} 
\rightarrow 0$ if $l_0 \neq 0$. Similarly, when $f \rightarrow \infty$,
~$Q_{\nabla T} \rightarrow 0$. The decrease of the heat flux at $f>1$ is 
simply due to the increase of the distance between the plates and 
corresponding decrease of the temperature gradient. The decrease at $f<1$ 
is due to systematic increase of the angle between the field lines and the 
direction of the temperature gradient.

\begin{figure}
\includegraphics[width=84mm]{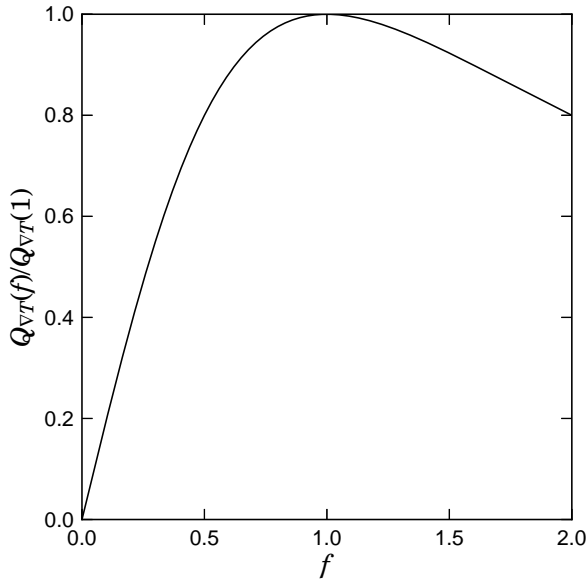}
\caption{Suppression of the heat flux along the temperature gradient 
between two approaching/receding plates as a function of distance $f$ 
between the plates, when the medium between the plates is threaded by 
tangled magnetic field [see equation~(\ref{eq:QgradT})]. At the initial 
moment ($f=1$), all angles between the magnetic field direction and the 
plates are equally probable. The decrease of the heat flux at $f>1$
is simply due to the increase of the distance between the plates and
corresponding decrease of the temperature gradient. The decrease at
$f<1$ is due to systematic increase of the angle between the field
lines and the direction of the temperature gradient.}
\label{fig:func}
\end{figure}

If at some moment the field lines are tangled in such a way that all angles
$\theta$ are equally probable, then parametrizing compression/stretching 
along $y$ by the same factor $f$ and averaging over $\theta$ gives us the 
suppressed heat flux along the temperature gradient:
\beq
\label{eq:QgradT}
Q_{\nabla T}=\chi \frac{T_2 - T_1}{h_0} \frac{2 f }{f^2+1}\\
(\rmn{see~Fig.}~\ref{fig:func}). 
\eeq

Thus, increasing the temperature gradient by squeezing the layers of the 
gas does not boost the heat exchange between them but rather makes it 
smaller. A qualitatively similar situation might occur at the cold fronts 
-- contact discontinuities formed by differential gas motions, a very 
simple model of which is discussed in the next subsection.

\subsection{Astrophysical example: model of a cold front}
\label{sec:cold}
{\it Chandra} observations of galaxy clusters often show sharp
discontinuities in the surface brightness of the ICM emission
\citep[see review by][]{MarkevichVikhl2007}. Most of these
structures have lower-temperature gas on the brighter (higher-density)
side of the discontinuity, suggesting that they are contact
discontinuities rather than shocks.  In the literature, these
structures are called `cold fronts'. Because of the sharp
temperature gradients, the limits on the thermal conduction derived for
the observed cold fronts are strong \citep[see e.g.][]{Ettori2000,
Vikhlinin2001, Xiang2007}.

\begin{figure}
\includegraphics[width=84mm]{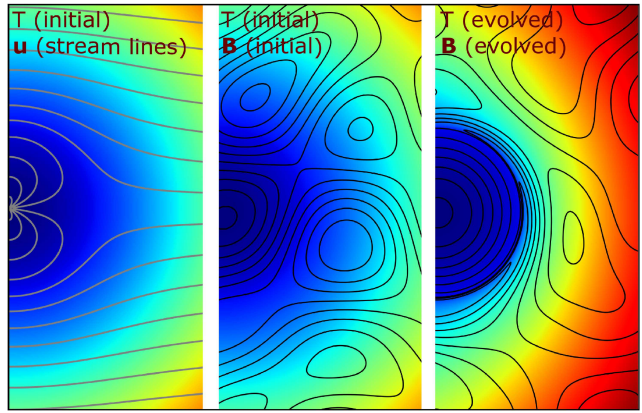}
\caption{Alignment of the field lines perpendicular to the
  temperature gradient for the velocity field characteristic of a
  cold front. A potential flow past a cylinder is
  used in this example. The left panel shows the initial temperature
  distribution (color image) and stream lines of the velocity
  field. The middle panel shows a random tangled magnetic field 
  superposed on the initial temperature distribution. The right panel 
  shows the time-evolved temperature map and magnetic-field lines 
  (superposed contours) in such a flow. The flow boosts the temperature 
  gradient at the cold front and at the same time stretches the field 
  lines along the lines of constant temperature. In the resulting 
  configuration, the field lines are essentially perpendicular to the
  sharp temperature gradient at the front.
}
\label{fig:cold_front}
\end{figure}

In the majority of theoretical models, the formation of a cold front involves
relative motion of cold and hot gases. Here we consider the
case of a hot gas flowing around a colder, gravitationally bound gas
cloud, which is a prototypical model of a cold front. For simplicity,
we assume that the velocity field can be approximated with a 2D
potential flow past a cylinder, while the initial temperature is
symmetric around the cylinder. The initial temperature distribution 
 and stream lines of the flow are shown in the left panel of
Fig.~\ref{fig:cold_front}. The middle panel shows the field lines of
a random magnetic field superimposed on the initial temperature
distribution. The evolved temperature and magnetic field are shown
in the right panel of Fig.~\ref{fig:cold_front}. Stretching of the
fluid elements near the stagnation point along the front leads to the
contraction of the same elements in the direction perpendicular to the
front. This configuration has been considered in a number of studies
of the cold fronts \citep[see e.g.][]{Asai2007, ChurazovInogamov2004,
Roediger2011, Lyutikov2006}. Qualitatively, it corresponds to the situation sketched
in Section~\ref{sec:example} and Fig.~\ref{fig:conv_flow}, which
naturally leads to the field lines orthogonal to the temperature
gradient at the front.

\subsection{Local correlation between the magnetic-field strength 
and the heat flux}
Let us now discuss the suppression of the local heat flux in more general 
terms. Consider the induction equation for an incompressible medium and 
the advection equation for the temperature:
\bea
\label{eq:B_adv}
\frac{\rmn{d} \vB}{\rmn{d} t} &=& \vB\cdot\nabla\vu,\\
\label{eq:temp_adv}
\frac{\rmn{d} T}{\rmn{d} t} &=& 0,
\eea
where $\vB$ is the magnetic field, $\vu$ the velocity field, 
$T$ temperature and $\rmn{d}/\rmn{d} t=\partial/\partial t+\vu\cdot\nabla$. 
We have neglected thermal and magnetic diffusivities. While 
equation~(\ref{eq:temp_adv}) is not a full magnetohydrodynamic energy equation, 
it is correct in the limit of an incompressible non-stratified medium (we will 
further discuss its applicability in Section~\ref{sec:disc}). Let $\vg$ be the unit vector 
in the direction of the temperature gradient, $\vb$ the unit vector in the 
direction of the field line, $B$ the magnetic field magnitude and $G$ the 
temperature gradient magnitude, so $\vB = B \vb$, $\nabla T = G \vg$. The above 
equations imply
\bea
\frac{\rmn{d}G}{\rmn{d}t} &=& - G \vg \cdot ( \nabla \vu) \cdot \vg,\\
\frac{\rmn{d}B}{\rmn{d}t} &=& B \vb \cdot (\nabla \vu) \cdot \vb,\\
\frac{\rmn{d} \mu}{\rmn{d}t} &=& \mu[\vg\cdot (\nabla \vu) \cdot \vg - 
\vb \cdot ( \nabla \vu) \cdot \vb],
\eea
where $\mu =\vb \cdot \vg$, the cosine of the angle between $\vB$ and $\nabla T$. 
From these equations, we can immediately infer the following equation for 
$\vb\cdot\nabla T = G \mu$, a quantity proportional to the parallel heat flux:
\beq \label{eq:1}
\frac{\rmn{d} \ln{(G \mu)}}{\rmn{d}t} = - \frac{\rmn{d} \ln{B}}{\rmn{d}t}.
\eeq
Thus, locally, the heat flux decreases as 
the field strength grows.

\subsection{Numerical example: a random 2D velocity field}
\label{sec:num2d}
\begin{figure}
\includegraphics[width=84mm]{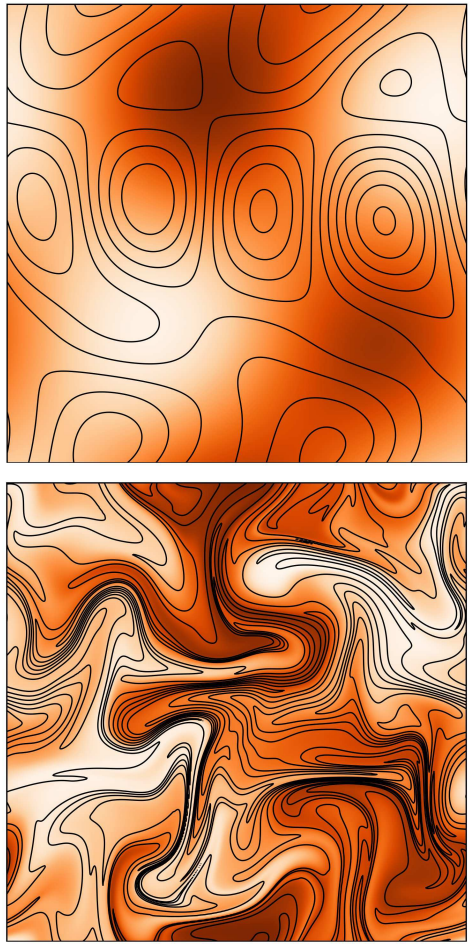}
\caption{Alignment of the field lines perpendicular to the temperature
  gradient for a stochastic $\delta$-correlated-in-time Gaussian incompressible 
  velocity field, modelled as a superposition of Fourier harmonics 
  with random phases and amplitudes. The top panel shows the initial random
  temperature distribution (color) with the field lines of a
  random magnetic field superposed (they are uncorrelated with temperature). 
  The bottom panel shows the same fields later on in the evolution. 
  In the evolved image, field lines follow the lines of constant temperature 
  in the regions where the temperature gradient is large.}
\label{fig:2d_2}
\end{figure}

In this example, we consider a random temperature distribution and
a random magnetic field in a random $\delta$-correlated-in-time (white) Gaussian 2D velocity field
(Fig.~\ref{fig:2d_2}). The temperature $T(x,y)$, the magnetic field 
 $\vc{B}(x,y)$ and the velocity field $\vc{u}(x,y)$ (assumed 
incompressible, $\nabla \cdot \vu=0$) are modelled as 
superpositions of Fourier harmonics with random phases and 
amplitudes. The temperature and the magnetic field are 
advected according to equations~(\ref{eq:B_adv}) 
and (\ref{eq:temp_adv}). The velocity field is renewed at each time 
step (white-in-time field). The initial conditions are shown in the top panel of
Fig.~\ref{fig:2d_2}; there is no initial correlation between
the temperature gradients and the orientation of the field lines. With
time, preferential stretching/squeezing of the fluid elements
leads to alignment of the field lines along the iso-temperature
lines (see bottom panel in Fig.~\ref{fig:2d_2}). This happens in all
regions where the stretching/squeezing is sufficiently strong. As a result,
the field lines are mostly perpendicular to the direction of the
temperature gradient in all regions where the gradient is large.
Intuitively, one expects that in a turbulent conducting medium, 
this tendency of local alignment between 
the magnetic filed and the isotherms will manifest itself statistically. In the next section, we work out 
a simple statistical model of this process. 
 
\section{Heat conduction in a stochastic velocity field}
\label{sec:results}
Here we treat the suppression of the heat conduction using an analytically 
solvable model that allows us to predict the statistical distribution 
of the cosine of the angle between the thermal gradient and the field line 
($\mu$), the magnitude of the thermal gradient ($G$) and the magnetic-field 
strength ($B$). After the joint probability distribution function (PDF) of 
$\mu$, $G$ and $B$ is derived (Section~\ref{sec:joint_pdf}), we will be in a position 
to assess how statistically prevalent the behaviour discussed in 
Section~\ref{sec:num2d} is, but we will preface this detailed calculation with 
some simpler arguments to quantify the suppression of the heat flux.

\subsection{Relaxation of temperature fluctuations}
\label{sec:wipe_time}
Let us restore heat conduction in equation~(\ref{eq:temp_adv}):
\beq
\label{eq:cond_eq}
\frac{\dl T}{\dl t} = \nabla \cdot (\chi \vb\vb\cdot\nabla T),
\eeq
where $\chi$ is the parallel thermal diffusivity coefficient 
\citep{Braginskii1965}. Then the volume-averaged rate of change 
of the rms temperature fluctuations is
\beq
\label{eq:temp_fluct}
\frac{\dl \langle \delta T^2 \rangle}{\dl t} = 
-2\chi\langle|\vb\cdot\nabla\delta T|^2\rangle=-2\chi\langle G^2\mu^2\rangle.
\eeq
Thus, the average value of $G^2 \mu^2$ characterizes the rate at which 
local temperature variations are wiped out by the thermal conduction.

\subsection{Kazantsev-Kraichnan model}
\label{sec:solv_model}
We consider the magnetic field to be so weak that it does not affect 
the velocity field. This condition
is only satisfied if the magnetic energy density is much lower than the
kinetic energy density of the plasma motions. This means that our model 
does not describe the saturated state, when these energy
densities become comparable. The non-saturated regime could be a common 
transient situation in the ICM, at least locally, in the sense that 
at any given time, the magnetic field is amplified up to the saturation 
value only in a small fraction of the volume. 

We will wish to calculate the joint PDF $p(\mu, G, B; t)$, where $\mu$ 
and $G$ are defined in Section~\ref{sec:qual_pic}, and investigate the
evolution of the relevant correlations, viz., $\langle G^2 \mu^2\rangle$ 
(see Section~\ref{sec:wipe_time}). To do that, we need to average the dynamical 
equations for $\vg$, $\vb$, $G$ and $B$ over all realizations of the 
stochastic velocity field. The equations are
\bea
\nonumber
\frac{\rmn{d}g^k}{\rmn{d}t} &=& - (\delta ^k_m - g^k g^m) g^i \partial_m u^i,\\
\nonumber
\frac{\rmn{d}b^k}{\rmn{d}t} &=&  (\delta ^k_i - b^k b^i) b^m \partial_m u^i,\\
\nonumber
\frac{\rmn{d}G}{\rmn{d}t} &=& - G g^i g^m \partial_m u^i,\\
\label{eq:gbG}
\frac{\dl B}{\dl t} &=& B b^i b^m \partial_m u^i,
\eea
where summation over repeated indices is implied.

This problem is solvable analytically for a Gaussian white-in-time
velocity field \citep{Kazantsev1968}:
\beq
\langle u^i(t, \vc{x}) u^j(t', \vc{x}') \rangle = 
\delta (t-t') \kappa^{ij} (\vc{x} - \vc{x}'),
\eeq
where $\kappa^{ij}$ is the correlation tensor, whose form can be determined
from symmetry and incompressibility considerations. We may assume the 
medium to be isotropic and homogeneous. Let us restrict our consideration to variation of magnetic field and temperature on spatial scales much smaller than that of the velocity field. Then, at any arbitrary point in space, the 
velocity can be expanded in linear approximation:
\beq
\label{eq:u_linexp}
u^i(t, \vc{x}) = \sigma ^i_m (t) x^m,
\eeq
where $\sigma^i_m (t) = \partial _m u^i$ and we have assumed $u^i(t,0)=0$ 
without loss of generality (otherwise change the reference frame). Then the 
velocity gradients satisfy
\bea
\nonumber
\label{eq:velcorr}
\left \langle \frac{\partial u^i}{\partial x^m}(t, \vc{x})  
\frac{\partial u^j}{\partial x'^n} (t', \vc{x}')\right \rangle 
\bigg\vert_{\bmath{x}=\bmath{x}'} &=& \langle \sigma^i_m(t) \sigma^j_n(t') 
\rangle\\
&=& \delta(t - t') \kappa^{ij}_{mn},
\eea
where
\[
\kappa^{ij}_{mn} = -\displaystyle \frac{\partial^2
\kappa^{ij}(\vc{y})}{\partial y_m \partial y_n}
\bigg\vert_{\bmath{y}=0} \equiv \kappa T^{ij}_{mn}
\]
and $\kappa = 1/ \tau_{eddy}$, $\tau_{eddy}$ being the turnover time of 
the turbulent eddies and 
\beq \label{eq:corr_tensor}
T^{ij}_{mn} = \delta^{ij}\delta_{mn} - \frac{1}{d+1} \lt 
(\delta^i_m \delta^j_n + \delta^i_n \delta^j_m \rt )
\eeq
is the inevitable tensor form of $\kappa^{ij}_{mn}$ for an isotropic 
incompressible medium of dimension $d$ ($=2,~3$). This is the so-called 
Kazantsev-Kraichnan model, which has been a popular tool for modelling 
the properties of small-scale dynamo and passive-scalar advection in 
turbulent media \citep[e.g.,][and references therein]{Chertkov1999, 
Balkovsky1999, Boldyrev2000, Schek2002small-sc, Schek2004passive, 
Boldyrev2004}.

\subsection{Relation between magnetic-field amplification and 
suppression of conduction for the white-in-time velocity field}
\label{sec:relationGmuB}
Before presenting the full statistical calculation, we wish to give a 
relatively simple one that establishes the connection between the 
relaxation rate of the temperature fluctuations and the magnetic-energy 
density. The heat flux along the field line $G \mu$ is inversely 
proportional to the length of a field-line segment $s$. Therefore, 
one can relate the change of the mean square heat flux $\langle G^2 \mu^2 
\rangle$, which is also the decay rate of the temperature fluctuations 
(see Section~\ref{sec:wipe_time}), to the growth of the magnetic-energy 
density as follows:
\beq
\label{eq:inv_prop}
\langle B^2 \rangle \propto \langle s^2 \rangle,~
\langle G^2 \mu^2 \rangle \propto \langle 1/s^2 \rangle.
\eeq
  
As explained in Section~\ref{sec:solv_model}, we assume an isotropic linear 
random velocity field. Let it be piecewise constant in time over intervals 
$\tau_c$ and completely uncorrelated for $\Delta t > \tau_c$. Assume further that
the amount of stretching of any fluid element over individual time intervals
of duration $\sim\tau_c$ is small compared to the size of the element, 
which amounts to a model of white-noise field. Under these assumptions, 
it is easy to obtain the PDF of $s$ as a function of time $t$ in the limit 
$t/\tau_c\gg 1$. The evolution of each component of the separation vector 
$\vc{x}$ of any two locations frozen into a velocity field constant over time 
interval $\tau_c$ is
\beq
x^i(\tau_c) \approx x^i(0)+\tau_c \sigma^i_j x^j(0)+\frac{1}{2}\tau_c^2 
\sigma^i_j \sigma^j_k x^k(0) + O(\tau_c^3),
\eeq
where $\sigma^i_j$ is the velocity gradients matrix 
[see equation~(\ref{eq:u_linexp})]. Since we are dealing with a random 
isotropic field, we can set $\vc{x}(0)=(1,0,0)$ at $t=0$. Then
\bea
\nonumber
\lefteqn{x^1(\tau_c) \approx 1 +\tau_c \sigma^1_1 + \frac{1}{2}\tau_c^2 
\sigma^1_j \sigma^j_1 + O(\tau_c^3),}&&\\
\label{eq:oneact}
\lefteqn{x^{i\neq1}(\tau_c) = \tau_c \sigma^i_1+O(\tau_c^2).}&&
\eea
We are interested in the time evolution of the `stretching factor' 
$s^2=|\vc{x}|^2$. For one `act of stretching', equation~(\ref{eq:oneact}) 
implies 
\beq
\ln{s^2(\tau_c)} = 2\tau_c \sigma^1_1 - 2\tau_c^2 (\sigma^1_1) ^2
 + \tau_c^2
\sigma^j_1 \sigma^j_1 +\tau_c^2 \sigma^1_j \sigma^j_1 +O(\tau_c^3).
\eeq
For $t \gg \tau_c$, the calculation of $s^2(t)$ reduces to summation of  
$N=t/\tau_c \gg 1$ such independent stretching episodes:
\beq
\lefteqn{
\ln{s^2(t)}=\sum{\ln{s^2(\tau_c)}}.
}
\eeq
After applying the central limit theorem to $(1/N) \sum \ln{s^2(\tau_c)}$, 
one readily gets the PDF of $s^2$:
\beq
P(s^2)=\frac{1}{s^2}\frac{1}{\sqrt{2\pi \sigma^2_{s}}} \exp\left[
-\frac{\left (
\ln{s^2}- m_{s} \right )^2}{2\sigma^2_{s}}\right], 
\eeq
where
\bea
\nonumber
\sigma_{s} &=& 2 \sqrt{T_{11}^{11}\frac{t}{\tau_{eddy}} },\\
m_{s} &=& \left [ - 2 T_{11}^{11}+ 
\sum_{i=1}^{d}\left ( T_{11}^{ii}+T_{1i}^{i1} \right )
 \right ] \frac{t}{\tau_{eddy}},
\eea
where $\tau_{eddy}$ and $T^{ij}_{mn}$ are defined at the end of 
Section~\ref{sec:solv_model}. We have taken $\delta(0)=1/\tau_c$ 
in equation~(\ref{eq:velcorr}). Using equation~(\ref{eq:inv_prop}), 
we get
\beq
\langle B^2 \rangle\propto e^{m_{s}+\sigma_{s}^2/2},~
\langle G^2\mu^2 \rangle \propto e^{-m_{s}+\sigma_{s}^2/2}.
\eeq
This leads to a simple relation between the growing magnetic-energy 
density and the evolution of the mean square heat flux:
\beq
\label{eq:GmuB}
\lefteqn{
\langle G^2\mu^2 \rangle \propto \langle B^2 \rangle^p,~\rmn{where}~
p=\frac{-m_{s}+\sigma_{s}^2/2}{~~m_{s}+\sigma_{s}^2/2}.
}
\eeq
For an incompressible velocity field in 3D, using equation~
(\ref{eq:corr_tensor}), we get $p=-1/5$. This is a statistical version 
of the dynamical equation~(\ref{eq:1}). It implies that on average, as 
the magnetic-energy density grows, the rate of decay of the temperature 
fluctuations is reduced, although the efficiency of this reduction is 
modest ($p$ is low). This is because $\langle G^2 \mu^2 \rangle$ is 
dominated by regions of low stretching while $\langle B^2 \rangle$ by 
regions of high stretching [equation~(\ref{eq:inv_prop})] and the 
distribution of these is highly intermittent.  

\subsection{Finite-time-correlated velocity field}
\label{sec:fin_corr}
How sensitive is this result to the obviously unphysical assumption of 
zero correlation time? Here, we numerically calculate the PDF of $s$ in a
random incompressible 3D velocity field evolving according to a Langevin 
equation with a finite correlation time. This is a generalization of the 
$\delta$-correlated case considered in Section~\ref{sec:relationGmuB}.

We consider a large number of independent field-line segments, each one 
placed in its own stochastic incompressible velocity field, given by 
equation~(\ref{eq:u_linexp}), with the velocity gradient satisfying
\beq
\label{eq:lang}
\frac{\rmn{d}\sigma^i_m}{\rmn{d}t} = -
\frac{1}{\tau_{c}} \sigma^i_m + \partial_m a^i,
\eeq
where $\tau_c$ is the correlation time and $a^i$ is a stochastic 
Gaussian acceleration whose gradient satisfies 
\beq 
\langle\partial_m a^i (t)\partial_n a^j(t')\rangle = 
\delta(t - t') A^2 T^{ij}_{mn}. 
\eeq 
Here $A^2$ is the noise amplitude and the dimensionless tensor $T^{ij}_{mn}$
is fixed by isotropy and incompressibility as given by 
equation~(\ref{eq:corr_tensor}). It is possible to define the effective 
turn-over time of turbulent eddies $\tau_{eddy}$ in much the same way 
as we we did for the $\delta$-correlated case: 
\beq 
\int _0^{\infty} \langle \sigma^i_m(0) 
\sigma^j_n (t) \rangle \rmn{d}t = \frac{1}{2} A^2
\tau_{c}^2 T^{ij}_{mn} \equiv \frac{1}{2 \tau_{eddy}}, 
\eeq 
where the exact solution of the Langevin equation (\ref{eq:lang}) has been 
substituted. Thus, $\tau_{eddy} = 1/(\tau_{c} A)^2$.

\begin{figure}
\includegraphics[width=84mm]{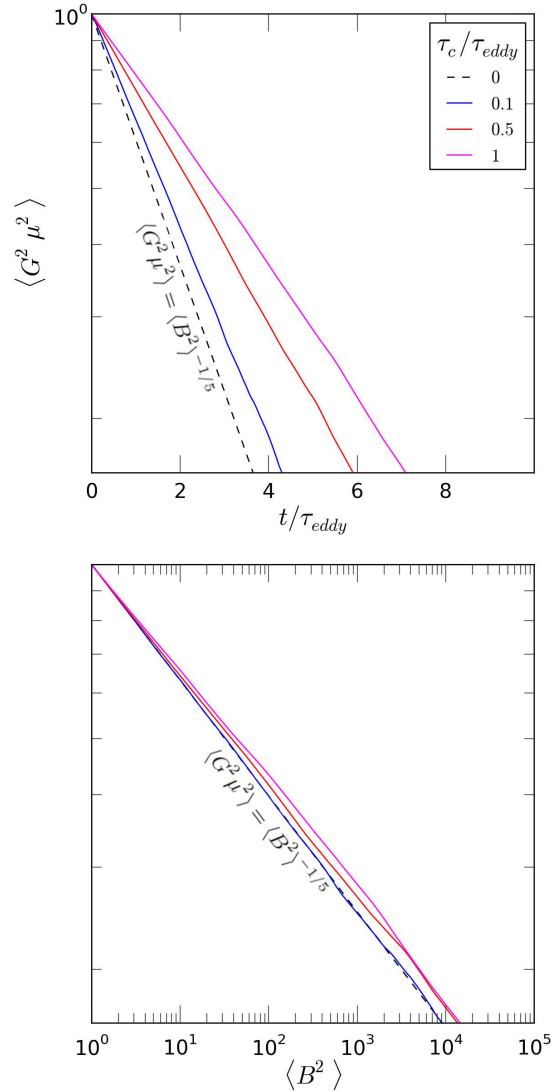}
\caption{The decrease of the mean square heat flux $\langle G^2\mu^2\rangle$ 
for the time-correlated velocity field and different ratios $\tau_c 
/ \tau_{eddy}$ (numerical results). While growth/decay rates of $\langle B^2 
\rangle$ and $\langle G^2 \mu^2 \rangle$ change with correlation time, their 
relative behavior is practically the same: $\langle G^2 \mu^2 \rangle \propto 
\langle B^2 \rangle ^{-0.2}$.}
\label{fig:non_corr} 
\end{figure}

In view of equation~(\ref{eq:inv_prop}), the evolution of 
$\langle G^2\mu^2\rangle$ and $\langle B^2\rangle$ can be easily calculated 
from the distribution of the segment lengths. Here we do this for a range 
of values of the ratio $\tau_{c} / \tau_{eddy}$. 
In Section~\ref{sec:relationGmuB}, we treated the case $\tau_{c} / \tau_{eddy} \rightarrow 0$ 
analytically, whereas for a physically sound case, $\tau_{c} / \tau_{eddy} \approx 1$ because 
typically turbulent velocities decorrelate over their eddy turnover times 
and fluid elements are stretched by order-unity amounts over the same 
time scales. The results are shown in Fig.~\ref{fig:non_corr}. Even though 
the growth/decay rates of $\langle B^2 \rangle$ and $\langle G^2 \mu^2 
\rangle$ do change with correlation time, their relative behaviour appears 
to be invariant, viz.,
\beq 
\langle G^2 \mu^2 \rangle \propto \langle B^2 \rangle ^{-1/5},
\eeq
practically the same as for the $\delta$-correlated regime [cf. 
equation~(\ref{eq:GmuB})]

Thus, finite correlation times do not change the form of the effective 
conduction-magnetic-energy-density relation, only modifying the time 
dependence. This result gives us some confidence in the Kazantsev-Kraichan 
velocity as a credible modelling choice.

\subsection{Statistics of the heat flux}
\label{sec:joint_pdf}

In this section we will finally derive the full joint statistical distribution 
of the fluctuating magnetic fields and temperature gradients and hence the detailed 
correlations between the heat flux, the field strength and the relative 
direction of the magnetic field and the temperature gradient.

For a velocity field given by equation~(\ref{eq:u_linexp}), we can
write equations~(\ref{eq:gbG}) for $\vg$, $\vb$, $G$ and $B$ as follows:
\bea
\label{eq:timeder}
\nonumber
\partial _t g^k &=& - (\delta ^k_m - g^k g^m) g^i \sigma ^i_m,\\
\nonumber
\partial _t b^k &=&  (\delta ^k_i - b^k b^i) b^m \sigma ^i_m,\\
\nonumber
\partial _t G &=& -G g^i g^m \sigma ^i_m,\\
\partial _t B &=& B b^i b^m \sigma ^i_m.
\eea
There are no advection terms here due to the homogeneity of the gas [so we 
can consider equation~(\ref{eq:gbG}) at $\vc{x}=0$].

The details of the derivation of the equation for the joint PDF $p(\mu,G,B;t)$ 
are presented in Appendix~\ref{app:A}. The result is
\bea
\label{eq:pdf}
\nonumber
\lefteqn{\partial _t p = \frac{\kappa}{2(d+1)} [ 2d(1 - \mu ^2)
 ( \mu \partial _{\mu} \mu \partial _{\mu} - \partial _G G \mu \partial _{\mu} 
- \partial_B B \mu \partial_{\mu}) }&&\\
\nonumber
&&+(d-1) (\partial _G G \partial _G G +\partial_B B \partial_B B) 
+ 2(1-\mu^2d)\partial_G G \partial_B B\\
\nonumber 
&&+ d(d + 1 - 2d \mu ^2) (2\mu \partial _{\mu} - \partial_G G - \partial_B B)\\ 
&&+ 2d^2(1 - d \mu^2) ] p,
\eea
where $d$ is the dimension of space. From now on, we only consider $d=3$.

Multiplying both sides of equation~(\ref{eq:pdf}) by $G^2\mu^2$ and 
integrating, we find
\beq
\label{eq:Gmu_mean_change}
\partial _t \langle G^2 \mu^2 \rangle = -\frac{{\kappa}}{2} 
\langle G^2 \mu^2 \rangle,
\eeq
so the mean square heat flux decays exponentially in time.
Then, recalling equation~(\ref{eq:temp_fluct}) for the rate of smoothing 
of the temperature fluctuations,
\beq
\frac{\dl \langle \delta T^2 \rangle}{\dl t} \propto - e^{-\kappa t/2} 
\rightarrow 0.
\eeq
We observe that the relaxation rate of the temperature
fluctuations decreases significantly on time-scales of the order of the
turnover time of the turbulent eddies ($\kappa = 1/\tau_{eddy}$).

It is also possible to recover the relation for the mean square heat flux 
as a function of the magnetic-energy density [equation~(\ref{eq:GmuB})]. 
Multiplying equation~(\ref{eq:pdf}) by $B^2$ and integrating, we obtain 
the evolution of the magnetic energy density:
\beq
\label{eq:evol_B2}
\partial _t \langle B^2 \rangle = \frac{5}{2}\kappa \langle B^2 \rangle.
\eeq
This result, combined with equation~(\ref{eq:Gmu_mean_change}), leads to 
the relation established in Section~\ref{sec:relationGmuB}:
\beq
\langle G^2 \mu^2 \rangle = \langle B^2 \rangle ^{-1/5}. 
\eeq

We expect that the temperature gradients and the magnetic-field lines will 
become perpendicular to each other. Let us then first investigate the limit 
of $\mu \rightarrow 0$, in which equation~(\ref{eq:pdf}) can be solved 
analytically. Let $x=\ln{\mu}$, $y=\ln{G}$ and $z=\ln{B}$. Then the joint PDF 
of these variables is $h(x,y,z;t) = p(\mu(x),G(y),B(z);t) e^{x+y+z}$, where 
the last factor is the Jacobian of the transformation of variables. Taking 
$\mu \rightarrow 0$ in equation~(\ref{eq:pdf}), we find that $h$ satisfies
\bea
\nonumber
\lefteqn{\partial_t h = \frac{\kappa}{4} 
[ 3 h_{xx}+ h_{yy}+h_{zz}-3(h_{xy}+h_{xz}) + h_{yz}}&&\\
\label{eq:pdf_h} 
&&+ 3(2h_x -h_y-h_z) ].
\eea
Let us now write $h$ in the following form:
\beq
h(x,y,z;t)=f(x,y;t)\delta(x+y+z).
\eeq
Substituting this ansatz into equation~(\ref{eq:pdf_h}), we find that the 
factorization goes through and $f$ satisfies
\beq
\label{eq:pdf_f}
\partial_t f = \frac{\kappa}{4} 
[ 3 f_{xx} + f_{yy} - 3f_{xy} + 3(2f_x -f_y) ].
\eeq
This factorization implies that in the limit $\mu\rightarrow 0$, 
$G \mu \propto 1/B$ independently of the initial conditions. This result 
was anticipated in Section~\ref{sec:relationGmuB}, where we took the ratio of 
$G \mu$ and $1/B$ to be the same for all the segments of the field lines at 
the initial moment.

Let us make another transformation: $\xi = x=\ln{\mu}$ and $\eta = x+2y = 
\ln{(G^2 \mu)}$ to separate variables in equation~(\ref{eq:pdf_f}). The joint 
PDF of these two variables, $w(\xi,\eta;t) = f(x(\xi),y(\xi,\eta);t)$, 
satisfies
\beq
\partial_t w = \frac{\kappa}{4} ( 3 w_{\xi \xi}+w_{\eta \eta}+6w_{\xi}).
\eeq
This equation can be readily solved:
\bea
\nonumber
\lefteqn{w(\xi,\eta;t) = \frac{1}{\sqrt{3}\pi\kappa t}\int_{-\infty}^{+\infty}\dl\xi'\dl\eta'
w(\xi',\eta';0) e^{-\frac{1}{3\kappa t}\lt[\frac{3}{2}\kappa t+(\xi-\xi')\rt]^2}}&&\\ 
&&\times e^{-\frac{1}{\kappa t}(\eta-\eta')^2 }.
\eea
Notice that along with diffusion in both variables, the PDF drifts to $\xi 
\rightarrow -\infty$, i.e to smaller $\mu$. So there is a continued tendency 
towards mutually perpendicular orientation of the thermal gradients and the 
field lines.

\begin{figure*}
\includegraphics[width=168mm]{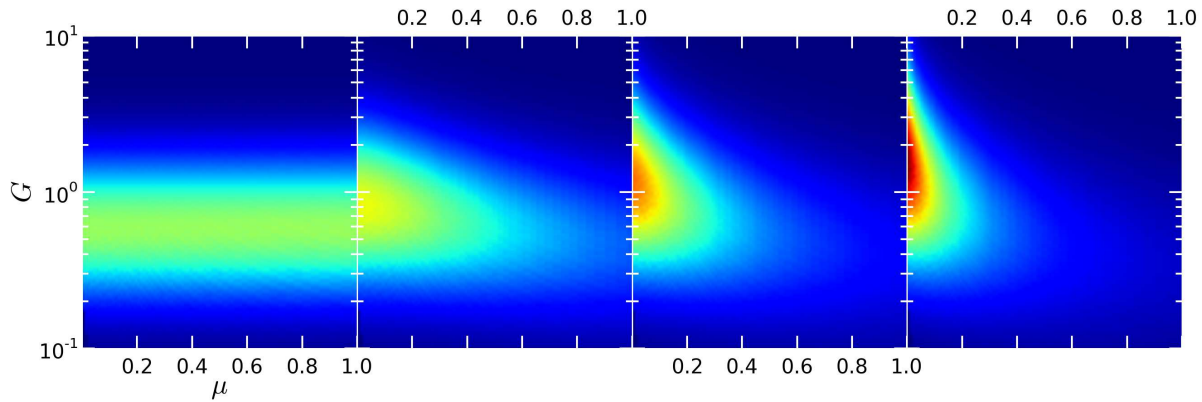}
\vbox to5mm{
\caption{Evolution of the joint PDF of $\mu$ and $G$ at 
regular time intervals from $t=0$ to $t=\tau_{eddy}$ (turn-over time of
the turbulent eddies) obtained via numerical solution of the 
equation~(\ref{eq:divpdf}). The maximum of the function drifts to the region 
where the thermal gradients and the field lines are perpendicular ($\mu
\rightarrow 0$).}
\label{fig:pdf}}
\vspace{5pt}
\end{figure*}

\begin{figure*}
\includegraphics[width=168mm]{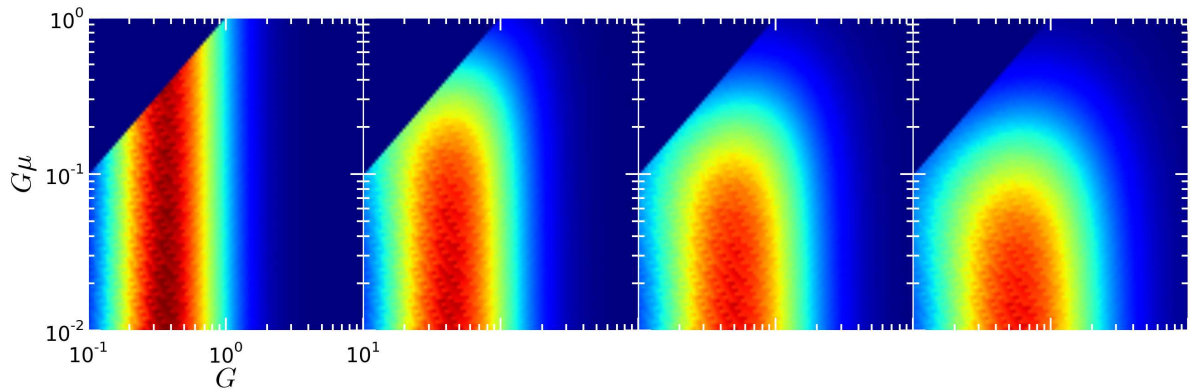}
\vbox to5mm{
\caption{Evolution of the joint PDF in terms of heat flux 
$G \mu = |\vc{b}\cdot\nabla T|$ and $G=|\nabla T|$ at the same times as in 
Fig.~\ref{fig:pdf}. Sharper gradients tend to be wiped out slower due to the 
smaller corresponding values of the heat flux.}
\label{fig:pdf_muG}}
\end{figure*}

If one is interested how the joint PDF of $\mu$ and $G$ behaves in the case 
of $\mu$ of order unity, the full equation~(\ref{eq:pdf}) integrated over $B$ has to 
be solved. Technically speaking, we are obliged to do 
this in order to ascertain that the limit $\mu\rightarrow 0$ was the relevant 
one to consider, i.e. that the joint distribution of $\mu$ and $G$ moves 
towards smaller $\mu$ independently of initial conditions. Again, to separate 
variables, we employ the variables $\xi=\ln{\mu}$ and $\eta=\ln{(G^2 \mu)}$. The 
PDF of these variables, $w(\xi,\eta;t)=\int p(\mu(\xi),G(\xi,\mu),B;t)
e^{\frac{1}{2}(\xi+\eta)}\dl B$, satisfies
\bea
\nonumber
\lefteqn{\partial _t w = \frac{{\kappa}}{4} [ 3(1 - e^{2\xi}) 
w_{\xi \xi} + (1 + 3e^{2 \xi})w_{\eta \eta}}&&\\
\label{eq:xieta}
&& + 6(1 - 2e^{2\xi})w_{\xi} - 12e^{2\xi}w ].
\eea
In order to solve this equation numerically, it is convenient to rewrite it 
in the divergence form as follows:
\bea
\label{eq:divpdf}
\nonumber
\lefteqn{\partial _t w = \frac{{\kappa}}{4} \{ \partial _{\xi} 
[ 2(1-e^{2\xi}) + (1-e^{2\xi})\partial _{\xi}]}&&\\
&&+\partial _{\eta}(1+3e^{2\xi}) \partial _{\eta} \} w.
\eea
Numerical solution of this equation is presented in Fig.~\ref{fig:pdf}. 
With time, the maximum of the PDF does indeed shift towards smaller $\mu$, 
demonstrating that the temperature gradient and the magnetic-field vector 
are becoming ever more orthogonal to each other. One can replot this 
graph in coordinates $G \mu$ (heat flux) and $G$ to observe that the rate of 
smearing of the temperature fluctuations in equation~(\ref{eq:temp_fluct}) is
correlated with the magnitude of the temperature gradients 
(Fig.~\ref{fig:pdf_muG}) in such a way that sharper gradients on average 
tend to be wiped out slower due to smaller corresponding values of $G \mu$.

\section{Limitations of our theory and a numerical test}
\label{sec:disc}
We now discuss the assumptions we have made in our theory and the 
extent to which they limit its applicability.
\subsection{Spatial scales}
The ordering of scales in the problem considered in this paper
obeys the following relations: 
\beq \rho_e\ll\lambda_{mfp}\lesssim l
\lesssim \lambda_u, 
\eeq 
where $l$ is the characteristic size of the region we deal with, $\rho_e$
is the electron Larmor radius, $\lambda_{mfp}$ is the electron mean-free 
path and $\lambda_u$ is the typical size of a turbulent eddy.   

The limit $l\ll\lambda_u$ simplifies the calculation of the field-line 
stretching because the linear expansion of the velocity field can
be used [equation~(\ref{eq:u_linexp})]. This allows for an analytic
treatment of the problem. Note that the kinematic dynamo naturally sets 
the parallel correlation length of the magnetic field $\lambda_{B\parallel}$ 
to be $\sim\lambda_u$ \citep{Schek2002small-sc, SchekCowleyTaylor2004}.

The condition $\lambda_{mpf}\lesssim l$ allows us to apply the thermal
conduction equation~(\ref{eq:cond_eq}) at these spatial scales. Due to the 
fact that in the kinematic-dynamo regime, $\lambda_u\sim\lambda_{B\parallel}$, 
we also have $\lambda_{mfp}\lesssim\lambda_{B\parallel}$. This limit being 
assumed, we can ignore the magnetic mirroring effects because the electrons 
are free to escape magnetic traps via collisional pitch-angle 
scattering \citep{CC1998, CC1999}.

The typical value of the mean-free path between Coulomb collisions, 
\begin{equation}
 \lambda_{mfp}\sim 8~ {\rm kpc} \left ( \frac{T}{5~{\rm KeV}} \right )^2 
 \left (\frac{n_e}{10^{-3}~{\rm cm^{-3}}} \right )^{-1},
\end{equation}
varies in cluster cores from $0.01$ to $20$ kpc depending on their 
temperature and density. For example, in the core of the Coma cluster, 
the mean-free path is $\sim 5$ kpc \citep{Churetal2012}; in M87/Virgo, 
it is much smaller, $\lambda_{mfp}\sim 0.01$ kpc, due to lower temperature 
and higher density \citep{ChurForman2008}. On the other hand, the value of 
$\lambda_u$ can be in the range of 10 kpc to 200 kpc \citep{Inogamov2003, 
Schuecker2004, Schek2006turb, Subramanian2006, Zhuravleva2011, KunzSchek2011}. 
Therefore, our analysis is relevant for temperature fluctuations on scales 
in the range $10^{-1} \mbox{--} 10^2$ kpc. Some of these scales are directly 
resolvable with {\it Chandra} or {\it XMM-Newton}, suggesting that in observed 
substructures in the temperature maps, the isotemperature contours should be 
roughly aligned with the magnetic-field lines.

\subsection{Incompressibility}
The assumption of incompressibility [needed to use equation~(\ref{eq:corr_tensor}] 
for the description of the velocity field) is valid as long as the gas velocities 
are subsonic. This is reasonable for the ICM, except for cases of strong mergers 
or AGN-driven strong shocks in the very core of a cluster. The comparison of 
cluster-mass estimates from X-ray data and lensing or stellar kinematics \citep[e.g.][]
{ChurForman2008} and simulations \citep[e.g.][]{LauKravtsov2009} suggest that the 
kinetic energy of the gas motions is at the level of 5-15\% of its thermal energy in 
relaxed clusters. Slight deviations from incompressibility should not dramatically alter 
our results.

\subsection{Stratification} 
We have neglected the effects of stratification. It is well known 
that in the ICM anisotropic thermal conduction modifies the classical 
Schwarzschild stability criterion in such a way that any radial temperature 
gradient leads to an instability: the magnetothermal instability (MTI, 
\citealt{Balbus2000}) if the thermal gradient and the gravity force are 
codirectional and the heat-flux-driven buoyancy instability (HBI, 
\citealt{Quataert2008}) if they are oppositely directed. These instabilities 
have been extensively studied in numerical simulations \citep{Sharma2009,
Parrish2009,Bogdanovic2009,RuszkowskiOh2010,Ruszkowski2011,McCourt2011,
KunzBogdan2012}. 

However, the instabilities are driven by large-scale mean gradients and in 
a turbulent plasma, at small enough scales, the buoyancy effects are likely 
to be less important than turbulent motions \citep[cf.][]{RuszkowskiOh2010} ---
essentially because turbulent time-scales get shorter at shorter spatial scales, 
while the buoyancy timescale is fixed. Indeed, the typical turbulent timescale 
is $\tau_{turb} \sim \lambda_u /u= \lambda_u/(M c_s)$, where $u$ is the 
(subsonic) velocity of turbulent motion, $M$ is the Mach number, $c_s$ is the 
speed of sound; in contrast, the buoyancy time-scale is $\tau_{buoy} \sim 
\sqrt{l_{P,T}/g} \sim \sqrt{l_{P,T}l_P}/c_s$, where $g$ is the gravitational 
acceleration caused by the cluster potential, $l_P$ is the characteristic length 
of change of the pressure profile and $l_T$ (in the case of MTI/HBI) is that of 
the macroscopic temperature profile. Therefore, $\tau_{turb} \lesssim 
\tau_{buoy}$ if $\lambda_u \lesssim M l_P$ or $\lambda_u \lesssim M 
\sqrt{l_P l_T}$ (for~MTI/HBI). Let $M$ be $\sim 0.3$. For typical cluster core 
parameters ($l_P\sim100$ kpc, $l_T\sim300$ kpc), we get $\lambda_u \lesssim 50$ 
kpc; for the bulk ($l_P\sim 300$ kpc, $l_T\sim 1000$ kpc), we get $\lambda_u 
\lesssim 200$ kpc. 

Thus, our results apply at smaller spatial scales, where the turbulence 
time-scales are shorter than the buoyancy time-scale. Obviously, one cannot 
neglect stratification while constructing a full self-consistent model of the 
ICM, but comparison with global cluster simulations (Section~\ref{sec:mhd_sim}) 
suggests that accounting for buoyancy does not eliminate the phenomenon of 
local orthogonalization of the field lines and temperature gradients.

\subsection{Thermal conduction}
Our model requires the eddy turnover time to be smaller than the 
conduction time, which is quite a serious restriction. Applying the 
standard formula for the Spitzer thermal conduction timescale, one gets
\bea
\label{eq:cond_est}
\nonumber
\lefteqn{\tau_{cond} \sim n_e k_B l^2 / \kappa_{Sp} \approx 3 \times 10^7 
\left ( \frac{n_e}{10^{-3} {\rm cm}^{-3}} \right ) 
\left (\frac{l}{100~{\rm kpc}} \right )^2 }&&\\
 && \times \left (\frac{T_e}{5~{\rm KeV}} \right )^{-5/2}~{\rm yr},
\eea
where $n_e$ is the electron density, $T_e$ the electron temperature and 
$\kappa_{Sp}$ the Spitzer thermal conductivity. At the same time,
\beq
\tau_{turb} \sim 5 \times 10^8~ \lt ( \frac{\lambda_u}{100~{\rm kpc}} \rt ) 
                                 \lt ( \frac{M}{0.3} \rt )^{-1}~{\rm yr}.
\eeq
From this estimate, it is clear that in the cool cores, the conduction 
time-scale can be longer than that of the turbulence, but in the hot 
($\sim 8~{\rm KeV}$) and rarefied ICM outside the core, the conduction 
time-scale can instead be much shorter. 

Nevertheless, even outside the core the orthogonalization of the temperature 
gradients and field lines is likely to take place. Qualitatively, this is 
because the effect of the orthogonalization is to switch off conduction, i.e. 
effectively lengthen the conduction time-scale compared to the estimate 
(\ref{eq:cond_est}). Thus, while gradients codirectional with field lines at some 
initial moment may be quickly erased by conduction, the ones that make large 
angles with the field lines can survive longer and, as turbulence orthogonalizes 
them further, conduction will become increasingly inefficient. In other words, 
the assumption of slow conduction will become better satisfied as the evolution 
proceeds. We will see that these qualitative arguments are indeed 
corroborated by cluster simulations with anisotropic conduction 
(\S~\ref{sec:mhd_sim}).  
 
\subsection{Dynamics of the magnetic field} 
As shown in Section~\ref{sec:relationGmuB}, the evolution of the decay rate
of the small-scale ($l\lesssim \lambda_u$) temperature
variations can be linked to the amount of stretching of the field
lines as $\propto \langle 1/s^2 \rangle$.  Essentially the decay rate
goes down because the field lines, along which the heat is transported, 
are stretched.\footnote{The effect of the field-line stretching on the
suppression of thermal conduction has previously been studied by
\citealt{RosnerTucker1989} and \citealt{Tao1995}, but in the case of
$\lambda_B < \lambda_{mfp}$ and constant macroscopic thermal gradient.}
The amount of stretching is, of course, limited by saturation of the magnetic 
field. This may turn out to be a key effect in the problem, but 
it is not analytically treatable as easily as the case of passive 
field considered here and is best addressed with direct numerical 
simulations (see Section~\ref{sec:mhd_sim}). Another potentially important 
effect we have ignored is the reconnection of the field lines, which may 
in principle considerably modify their topology. While we believe the simple 
model considered in this paper correctly captures the qualitative picture, 
direct numerical simulations are required to confirm this.

\subsection{Local versus global conduction} 
We stress again that we have only considered the suppression of \textit{local} 
thermal conduction, as applies to temperature \textit{fluctuations} on scales 
$l < \lambda_u$. We have established that the gradients associated with these 
fluctuations are predominantly oriented perpendicular to the magnetic-field lines 
by the plasma flow. In general, however, if one is interested in the global heat 
transport on scales $l\gg\lambda_u$, other effects start to be important: in 
particular, exponential divergence of field lines \citep{RR1978, Narayan2001, 
ChandranMaron2004}. 

\subsection{Comparison with global cluster simulations}
\label{sec:mhd_sim}
To back up our qualitative arguments in support of the conclusions of our 
theoretical model despite the many simplifying assumptions that were required 
to make it solvable, we have employed the data drawn from the simulations 
reported by \citet{ZuHone2013}. These are global dynamic MHD simulations of 
a disturbed cluster, which were not specifically designed to test our model but 
represent a current state-of-the-art numerical model of cluster evolution in response 
to a minor merger. The simulations incorporate all of the additional physics that 
we neglected and that is essential for a global model: a range of spatial and 
time-scales, compressibility, large-scale stratification, buoyancy, anisotropic thermal 
conduction, radiative cooling and dynamic back-reaction of the magnetic field on 
the fluid motions.

\begin{figure}
\includegraphics[width=84mm]{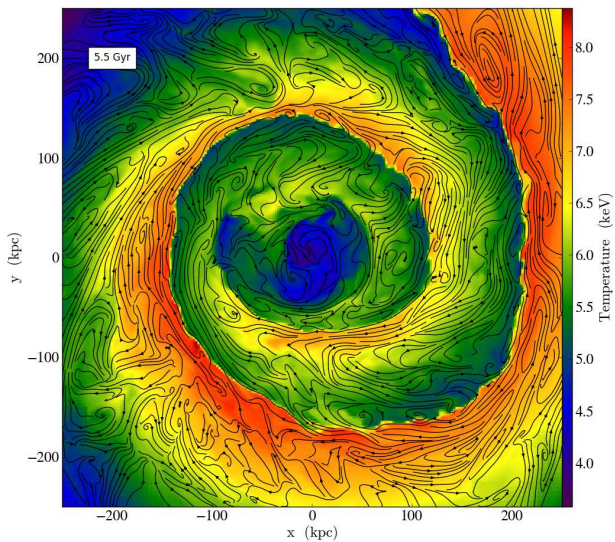}
\caption{Central 500-kpc region of the simulation box of \citet{ZuHone2013} at 
$t=5.5~{\rm Gyr}$ used for comparison of our theory with the simulations.}
\label{fig:zuhone_box}
\end{figure} 

In these simulations, a massive ($M \approx 1.5 \times 10^{15}~M_{\odot}$, $T\sim8$ keV), 
cool-core cluster, initially in hydrostatic equilibrium, merged with a small 
(mass ratio $R=5$) gasless subcluster, which set off the sloshing of the cool 
core. The simulation started at a point in time when the cluster centers had a 
mutual separation of $d = 3$ Mpc and an impact parameter $b = 500$ kpc. The
initial velocities of the subclusters were set up assuming that the total kinetic 
energy of the system was set to half of its total potential energy. The main 
cluster was set up within a cubical computational domain of width $L = 2.4$ Mpc 
on a side, with the finest cell size on the grid of 2.34 kpc. A random magnetic 
field was set up in Fourier space using independent normal random deviates for 
the real and imaginary components of the field. The field spectrum corresponded 
to a Kolmogorov shape with cut-offs at large ($\approx 500$ kpc) and small 
($\approx 40$ kpc) linear scales. The initial plasma $\beta$ was 400. 
For the detailed description of these simulations, see \citet{ZuHone2013} 
and \citet{ZuHone2011}. 
In the tests reported below, we only considered the central 500 kpc of the 
simulated cluster (Fig.~\ref{fig:zuhone_box}), where the disturbance of the ICM 
was greatest, leading to significant local temperature variation and tangled 
magnetic field.

\begin{figure*}
\includegraphics[width=168mm]{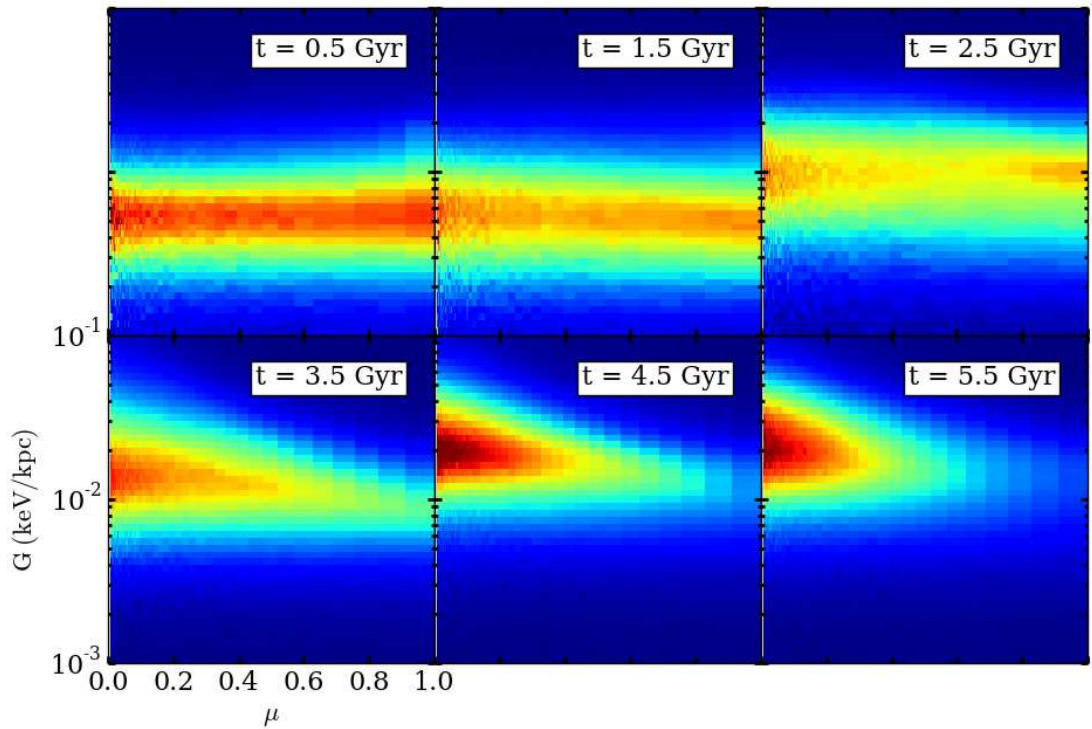}
\caption{Evolution of the joint PDF of $\mu = |\vc{b}\cdot\nabla T|$ and 
$G=|\nabla T|$ for the Run S1 of \citet{ZuHone2013} -- 
a global MHD cluster simulation with thermal conduction 
and radiative cooling switched off.}
\label{fig:zuhone_imhd}
\end{figure*}

First, we examined a simulation with thermal conduction and radiative cooling turned off 
-- this is the Run S1 from \citet{ZuHone2013}. Fig.~\ref{fig:zuhone_imhd} shows 
the evolution of the joint PDF$(G,\mu)$ analogous to that shown in Fig.~\ref{fig:pdf}. 
Initially, the magnetic field orientation was random as indicated by the flat PDF over 
$\mu$ in the top left frame in Fig.~\ref{fig:zuhone_imhd}. With time, the most probable 
value of $G$ increased, while the corresponding $\mu$ decreased. This behavior is 
qualitatively very similar to the evolution of the model PDF shown in Fig.~\ref{fig:pdf}. 
Note that in Fig.~\ref{fig:zuhone_box}, it is visually manifest that magnetic field 
lines and surfaces of constant temperature are aligned in much of the disturbed ICM, 
including both sharp fronts (cf. Section~\ref{sec:cold}) and the more random turbulent regions.

When radiative cooling was switched on, while conduction was still off (Run SX of 
\citealt{ZuHone2013}), the behaviour of the joint PDF of $G$ and $\mu$ was qualitatively 
very similar to that in pure MHD case described above, except for a small increase 
of the PDF at higher gradients independent of the value of $\mu$, which is expected because 
cooling may generate temperature gradients from the density gradients regardless of the 
magnetic-field orientation.

\begin{figure}
\includegraphics[width=84mm]{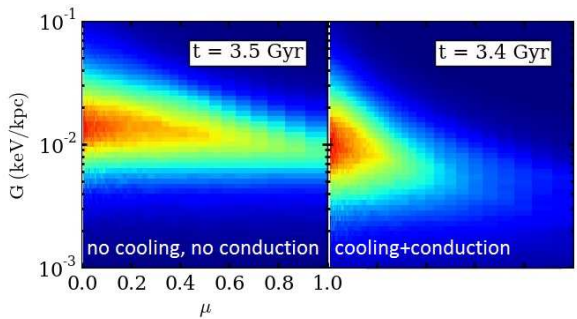}
\caption{Joint PDF of $G$ and $\mu$ in a global cluster simulations. 
Left panel: same case as in Fig.~\ref{fig:zuhone_imhd}. Right panel: 
identical case except with anisotropic conduction and radiative cooling 
switched on (Run SCX1 of \citealt{ZuHone2013}).}
\label{fig:zuhone_mhd_cond}
\end{figure}

Finally, consider a simulation identical to the ones used above but with both cooling and 
anisotropic conduction switched on (Run SCX1 of \citealt{ZuHone2013}). In Fig.~
\ref{fig:zuhone_mhd_cond}, the joint PDF of $G$ and $\mu$ for the cluster, taken at a late 
stage in its evolution, is contrasted to the case without conduction at a similar time. For 
this hot ($T\sim8 ~{\rm KeV}$) simulated cluster, thermal conduction was strong enough to 
make a non-trivial impact on the PDF. Efficient anisotropic conduction quickly eliminated 
small-scale thermal gradients in the regions where the field lines and the temperature 
gradients were aligned in the initial setup, while the gradients orthogonal to the field 
survived longer. This 
process shifted the PDF to smaller $\mu$ and lower $G$ as compared to the case with no 
conduction (see Fig.~\ref{fig:zuhone_imhd}). At the same time, the high gradients in the 
regions with small $\mu$ were preserved and enhanced on average by the gas motions. The 
maximum of the PDF still drifted to higher gradients and smaller $\mu$ so that high 
temperature gradients ended up associated with perpendicular orientation of the gradients 
and the field lines. Clearly, both effects (one driven by gas motions and another by 
anisotropic conduction) lead to a similar net result: large temperature gradients one can 
expect to find in the ICM are likely associated with regions where $\mu$ is small. Thus, the 
shape of the PDF derived from the simulation with cooling and thermal conduction is 
qualitatively similar to the one in the absence of cooling and thermal conduction (see Fig.~
\ref{fig:zuhone_mhd_cond}). We conclude that the effect proposed in this paper is identifiable 
even if efficient thermal conduction smears out the initial gradients on small scales.

Finally, let us stress again that these simulations were not specifically tailored for the 
problem at hand. For example, in our theoretical model, we assume continuous and spatially 
homogeneous driving with a well defined eddy turn-over time scale, while in the simulations, 
the cluster is perturbed at a specific time and in a special way. Also, on the numerical side, 
a precise evaluation of the angle between the field lines and gradients in the presence of 
small-scale eddies should have considerable uncertainty, precluding firm conclusion on the 
behavior of the PDF at very small $\mu$. This makes further detailed quantitative comparisons 
between theory and simulations problematic. A bespoke numerical study would clearly be a 
worthwhile undertaking and is left for the future. However, it appears that even on the basis 
of this limited comparison, we can conclude optimistically that the correlation between large 
gradients of $T$ and small values of $\mu$ is clearly present in the numerical model, which 
does not suffer from the limitations of our theory and contains much of the physics currently 
believed to be relevant.\footnote{Although it still takes no account of some of the plasma 
microphysics whose role remains poorly understood if potentially dramatic \citep{KunzSchek2011,
MogaveroSchek}.}

\section{Conclusions}
\label{sec:concl}
We have studied the correlations between the local fluctuating temperature 
gradients and the orientation of the frozen-in magnetic-field lines in the 
turbulent ICM. We have argued that the mutual orientation between isotherms 
and magnetic-field lines is not random, but rather a strong alignment is 
expected: gas motions tend to increase the temperature gradients and, at the 
same time, align the field lines perpendicular to the gradients. Cold 
fronts in clusters provide a vivid example of this process on large scales. 
The net result of the correlated evolution of the temperature distribution 
and the magnetic field is the effective suppression of the local heat flux. 
We note that global thermal conduction defined by radial temperature 
profiles of galaxy clusters (wich is one of the possible solutions to 
the cooling flow problem) is beyond the scope of this work.

We have calculated explicitly the joint distribution function of the gradients 
and the angles they make relative to the field lines and demonstrated that 
significant suppression takes place for generic 3D isotropic incompressible motions. 
The main conclusions are as follows: 
\begin{itemize}
  \item Strong correlation of the fluctuating temperature gradients and the local 
  magnetic field orientation is established on the timescale of the turbulent eddy 
  turnover.
  \item On average, the decay rate of temperature fluctuations is anti-correlated with 
  the degree of amplification of the magnetic field by the gas motions. Volume averaged 
  decay rate decreases with the growth of the magnetic-energy density as $\langle B^2 
  \rangle^{-1/5}$.\\
  
For disturbed clusters, where large-scale clumps of gas are displaced, the largest 
observed \textit{local} gradients should be associated with the largest heat flux 
suppression. The estimates of the effective conductivity based on these gradients may 
not be characteristic of the bulk of the gas. This conclusion appears to be supported 
by global dynamic cluster simulations with and without anisotropic conduction.    
\end{itemize}

\section*{Acknowledgements}
This work was supported in part by the Leverhulme Trust Network on Magnetized 
Plasma Turbulence. 

\bibliographystyle{mn2e}
\bibliography{bibliography}

\appendix

\section{Statistical calculation of the joint PDF of $\mu$, $G$ and $B$}
\label{app:A}
The general form of the joint PDF of the magnetic field and the temperature 
gradient is
\bea
\nonumber
\lefteqn{P(\vg, \vb, G, B; t)  = \langle\tilde{P}\rangle,}&&\\
\lefteqn{\tilde{P} = \delta (\vg - \vg(t)) 
                     \delta (\vb - \vb(t)) 
                     \delta (G - G(t))
                     \delta (B-B(t)),}&&
\eea
where $\vg$, $\vb$, $G$ and $B$ are variables and $\vg(t)$, $\vb(t)$, 
$G(t)$ and $B(t)$ are stochastic processes that are solutions of 
equations~(\ref{eq:timeder}). Taking time derivative of $\tilde{P}$ and 
using equations~(\ref{eq:timeder}), we obtain
\beq
\label{eq:operl}
\partial_t P = \rmn{\hat{L}}^m_i \sigma ^i_m \tilde{P},
\eeq
where
\bea
\nonumber
\lefteqn{ \hat{\rmn{L}}^m_i = \frac{\partial}{\partial g^k} 
        (\delta ^k_m - g^k g^m) g^i - \frac{\partial}{\partial b^k} 
        (\delta^k_i - b^k b^i) b^m}&&\\  
&&+ \frac{\partial}{\partial G} g^i g^m G - 
    \frac{\partial}{\partial B} b^i b^m B. 
\eea
The average of equation~(\ref{eq:operl}) is
\beq
\label{eq:dPdt}
\partial_t P = \rmn{\hat{L}}^m_i \langle \sigma ^i_m \tilde{P} \rangle
\eeq
and we now apply the Furutsu-Novikov formula \citep{Furutsu,Novikov} 
to calculate the right-hand side: 
\bea
\label{eq:aver}
\nonumber
\lefteqn{ \langle \sigma ^i_m(t) \tilde{P}(t)\rangle = 
    \int \rmn{d}t' \langle \sigma ^i_m(t) \sigma ^j_n(t') \rangle 
    \bigg \langle \frac{\delta \tilde{P}(t)}{\delta \sigma ^j_n(t')} 
                                                 \bigg \rangle }&&\\
    && = \kappa T^{ij}_{mn} 
    \bigg \langle \frac{\delta \tilde{P}(t)}{\delta \sigma ^j_n(t)} 
                                                  \bigg \rangle
\eea
where we have used equation~(\ref{eq:velcorr}). 
From equation~(\ref{eq:operl}),
\bea
\label{eq:var_der}
\nonumber
\lefteqn{ \frac{\delta \tilde{P}(t)}{\delta \sigma ^j_n (t)} = 
    \int_{-\infty}^t \rmn{d} t' 
    \bigg[ \rmn{\hat{L}}^m_i \delta ^i_j \delta ^n_m \delta(t - t') 
                                         \tilde{P} (t') }&&\\
    &&+\rmn{\hat{L}}^m_i \sigma ^i_m (t') 
    \frac{\delta \tilde{P} (t')}{\delta \sigma ^j_n (t)} \bigg] 
    = \frac{1}{2} \rmn{\hat{L}}^n_j \tilde{P}(t).
\eea
The second term inside the integral vanishes by causality ($t'<t$). 
Using equation~(\ref{eq:var_der}) in equation~(\ref{eq:aver}) and 
substituting into equation~(\ref{eq:dPdt}), we arrive at a closed equation 
for the desired PDF:
\beq \label{eq:eqinoperform}
\partial_t P = \frac{\kappa}{2} T^{ij}_{mn} \rmn{\hat{L}}^m_i 
                                              \rmn{\hat{L}}^n_j P.
\eeq

Since the medium is isotropic, the PDF only depends on $G$, $B$ and the 
angle between the unit vectors $\vg$ and $\vb$. Therefore, it can be 
factorized as
\beq
P(\vg, \vb, G, B; t) = \frac{1}{8\pi^2}\delta (\vg ^2 - 1) \delta (\vb ^2 - 1) 
p(\mu, G, B; t),
\eeq
where $\mu=\vc{b}\cdot\vc{g}$. The factor $1/8\pi^2$ has been introduced in 
order to keep $p(\mu,G,B;t)$ normalized to unity. Substituting this 
expression into equation~(\ref{eq:eqinoperform}), we get 
\bea
\label{eq:superlong}
\nonumber
\lefteqn{ \rmn{\hat{L}}^m_i \rmn{\hat{L}}^n_j P 
= \delta (\vg ^2 - 1) \delta (\vb ^2 - 1) }&&\\
\nonumber
&& \times \lbrace ( b^i b^j b^m b^n  + g^i g^j g^m g^n - g^i b^j g^m b^n  \\
\nonumber
&&- b^i g^j b^m g^n) \mu \partial_{\mu} \mu \partial_{\mu} \\
\nonumber
&&+ (b^i g^j b^m g^n - g^i g^j g^m g^n ) \mu \partial_{\mu} \partial_G G \\
\nonumber
&&+ (g^i b^j g^m b^n  - 2 b^i b^j b^m b^n + b^i g^j b^m g^n) 
\mu \partial_{\mu} \partial_B B \\
\nonumber
&&- ( g^i b^j g^m b^n + b^i g^j b^m g^n )\partial_G G \partial_B B \\
\nonumber
&&+ g^i g^j g^m g^n \partial_G G \partial_G G \\
\nonumber
&&+ [ 2(d+1)(b^i b^j b^m b^n + g^i g^j g^m g^n) \\
\nonumber
&&- 2d( g^i b^j g^m b^n + b^i g^j b^m g^n) - b^m b^n \delta ^i_j - 
b^j b^m \delta ^i_n \\
\nonumber
 &&- g^i g^n \delta ^j_m - g^i g^j \delta ^m_n ] \mu \partial_{\mu} \\
\nonumber
 &&+ [ -2(d+1)g^i g^j g^m g^n + d(g^i b^j g^m b^n +  b^i g^j b^m g^n) \\
\nonumber
 &&+ g^i g^n \delta ^j_m + g^i g^j \delta ^m_n ] \partial_G G \\
\nonumber
 &&+ [ -2(d+1)b^i b^j b^m b^n + d g^i b^j g^m b^n  + d b^i g^j b^m g^n \\
\nonumber 
 &&+ b^mb^n\delta^i_j + b^j b^m \delta^j_m  ] \partial_B B \\
\nonumber
 &&+ d [ (d+2)(b^i b^j b^m b^n + g^i g^j g^m g^n) - d(g^i b^j g^m b^n  \\
\nonumber
 &&+ b^i g^j b^m g^n) \\
 &&- (b^m b^n \delta ^i_j + b^j b^m \delta ^i_n + g^i g^n \delta ^j_m + 
g^i g^j \delta ^m_n) ] \rbrace  p, 
\eea
where $d$ is the number of spatial dimensions.
The PDF is factorized, as it ought to be, and we only need to solve the 
equation for $p(\mu, G, B; t)$. Substituting equation~(\ref{eq:superlong}) 
into equation~(\ref{eq:eqinoperform}), we perform the convolutions 
involving $T^{ij}_{mn}$ [see equation~(\ref{eq:corr_tensor})] using the 
identities
\bea
\nonumber
\lefteqn{T^{ij}_{mn} b^i b^j b^m b^n = \frac{d - 1}{d + 1},}&&\\
\nonumber
\lefteqn{T^{ij}_{mn} g^i b^j g^m b^n = \frac{\mu ^2 - 1}{d+1},}&&\\
\nonumber
\lefteqn{T^{ij}_{mn} b^j b^m \delta ^i_n = 0,}&&\\
\nonumber
\lefteqn{T^{ij}_{mn} b^m b^n \delta ^i_j = \frac{(d-1)(d+2)}{d+1},}&&\\
\nonumber
\lefteqn{T^{ij}_{mn} b^i g^j b^m g^n = \frac{\mu ^2 - 1}{d+1},}&&\\
\nonumber
\lefteqn{T^{ij}_{mn} g^i g^j g^m g^n = \frac{d-1}{d+1},}&&\\
\nonumber
\lefteqn{T^{ij}_{mn} g^i g^n \delta ^j_m = 0,}&&\\
\lefteqn{T^{ij}_{mn} g^i g^j \delta ^m_n = \frac{(d-1)(d-2)}{d+1}.}&&                 
\eea
The result is equation~(\ref{eq:pdf}).

\label{lastpage}

\end{document}